\documentclass[aps]{revtex4-1}
\usepackage[utf8]{inputenc}
\usepackage[T1]{fontenc}
\usepackage[english]{babel}
\usepackage{amssymb,amsmath,mathtools} 
\usepackage{graphicx} 
\usepackage{xcolor}
\usepackage{bbold}
\usepackage{braket}
\usepackage[headings]{fullpage}
\usepackage{hyperref}
\usepackage{slashed}
\usepackage[titletoc,toc,title]{appendix}
\usepackage[hang,small,bf]{caption}
\usepackage[nottoc,numbib]{tocbibind}
\usepackage[export]{adjustbox}
\usepackage{xcolor}
\usepackage{array}
\usepackage{booktabs}
\usepackage{hyperref}
\usepackage{amsthm}
\usepackage{lineno}
\usepackage{rotating}

\begin{document}
	\newcommand{\be}{\begin{eqnarray}}
		\newcommand{\ee}{\end{eqnarray}}
	\newcommand{\del}{\partial}
	\newcommand{\nn}{\nonumber}
	\newcommand{\STr}{{\rm Str}}
	\newcommand{\Sdet}{{\rm Sdet}}
	\newcommand{\Pf}{{\rm Pf}}
	\newcommand{\mat}{\left ( \begin{array}{cc}}
		\newcommand{\emat}{\end{array} \right )}
	\newcommand{\vect}{\left ( \begin{array}{c}}
		\newcommand{\evect}{\end{array} \right )}
	\newcommand{\tr}{{\rm Tr}}
	\newcommand{\hm}{\hat m}
	\newcommand{\ha}{\hat a}
	\newcommand{\hz}{\hat z}
	\newcommand{\hze}{\hat \zeta}
	\newcommand{\hx}{\hat x}
	\newcommand{\hy}{\hat y}
	\newcommand{\tm}{\tilde{m}}
	\newcommand{\ta}{\tilde{a}}
	\newcommand{\U}{\rm U}
	\newcommand{\D}{\slashed{D}}
	\newcommand{\hc}{^\dagger}
	\newcommand{\inv}{^{-1}}
	\newcommand{\diag}{{\rm diag}}
	\newcommand{\sign}{{\rm sign}}
	\newcommand{\ct}{\tilde{c}}
	\newtheorem{theorem}{Conjecture}
	\newcommand{\eins}{\leavevmode\hbox{\small1\kern-3.8pt\normalsize1}}
	
	\title{Memory Effects in Disease Modelling Through\\Kernel Estimates with Oscillatory Time History}

	\author{Adam Mielke}\email{admi@dtu.dk}\affiliation{Dynamical Systems, Department of Applied Mathematics and Computer Science, Technical University of Denmark, Asmussens Allé, 303B, 2800 Kgs.\ Lyngby, Denmark}
	\author{Mads Peter Sørensen}\email{mpso@dtu.dk}\affiliation{Dynamical Systems, Department of Applied Mathematics and Computer Science, Technical University of Denmark, Asmussens Allé, 303B, 2800 Kgs.\ Lyngby, Denmark}
	\author{John Wyller}\email{john.wyller@nmbu.no}\affiliation{Department of Mathematics, Faculty of Science and Technology, Norwegian University of Life Sciences, P.O. Box 5003, NO-1432 Ås, Norway}
	
	\date{\today}
    
	\begin{abstract}
	We design a linear chain trick algorithm for dynamical systems for which we have oscillatory time histories in the distributed time delay. We make use of this algorithmic framework to analyse memory effects in disease evolution in a population. The modelling is based on a susceptible-infected-recovered SIR - model and on a susceptible-exposed-infected-recovered SEIR - model through a kernel that dampens the activity based on the recent history of infectious individuals. This corresponds to adaptive behavior in the population or through governmental non-pharmaceutical interventions. We use the linear chain trick to show that such a model may be written in a Markovian way, and we analyze the stability of the system. We find that the adaptive behavior gives rise to either a stable equilibrium point or a stable limit cycle for a close to constant number of susceptibles, i.e.\ locally in time. We also show that the attack rate for this model is lower than it would be without the dampening, although the adaptive behavior disappears as time goes to infinity and the number of infected goes to zero.
	\end{abstract}
	
	\date{\today}
	\maketitle

\section{Introduction}\label{sec1}

Memory effects are an important part of disease modeling \cite{roddam2001mathematical, Memory1, Memory2, Memory3, Memory4, Memory5, Memory6}. Non-pharmaceutical interventions and individuals adapting their behavior in response to the news both fall under this category, and the challenge when implementing these in models is the non-local interaction in time, i.e., that the information about number of infected is delayed.
\\
\newline
In this paper, we apply kernel methods from animal population dynamics to epidemiological models as the susceptible-infected-recovered SIR - model and the susceptible-exposed-infected-recovered SEIR - model. These models allow analytical calculation of equilibrium points and their respective stability properties. We follow a modeling tradition which is common in theoretical ecology based on a dynamical systems approach with a distributed time delay incorporated, see \cite{MurrayVol12002}, \cite{MurrayVol22002},  \cite{deRoos2014} and \cite{cushing2013} and the references therein. A notable feature is that the distributed time delay has the significant advantage of being Markovian. Especially the criterion for an outbreak, local stability in time (i.e., for approximately constant number of susceptibles), and the attack rate are of interest. It is known that using non-pharmaceutical interventions such as lockdowns allows control of the system \cite{SIR-control}, but we here show that stability comes automatically from adaptive behavior.\\
\newline
The paper is divided into two main parts: First, we investigate the properties of SIR- and SEIR-models with added feedback on the activity based on the recent history of the number of infected. Second, we design a linear chain trick algorithm for a general dynamical delay system with oscillations in its history. For the sake of completeness we also show in detail the continuous dependence of the solutions of the dynamical systems in the time histories. One important outcome of this analysis is the robustness property of the solutions. Technical details can be found in Appendix \ref{Sec:AttackRateApp}

\section{SEIR-model with Memory Effects and its Basic Properties}\label{Sec:SEIR}

Let us start with the standard stratified SEIR model \cite{SIR1, SIR2, SIR3, SIR4, SIR5, arino2005multi} with $n$ groups in full generality, where $S,E,I,R: (0, \infty) \rightarrow \mathbb{R}^n$. Here $S$ denotes the number of susceptible, $E$ the number of exposed, $I$ the number of infectious and $R$ denotes the number of recovered individuals. These functions depend on time $t \in (0,\infty)$. The coordinates of $S,E,I,R$ counts the group members in the stratification, which may result from a segmentation of the population according to age, occupation, education and others. In addition we may include spatial segmentation  into countries, provinces, regions, cities and towns of the population. This gives us the following model 
\begin{equation}\label{Eq:SEIR-model}
	\left.\begin{aligned}
		\dot{S}\ =&\ - \diag(S)\beta I\\
		\dot{E}\ =&\   \diag(S) \beta I - \eta E\\
		\dot{I}\ =&\ \eta E - \gamma I\\
		\dot{R}\ =&\ \gamma I
	\end{aligned}\right.\
\end{equation}
where $\diag(S)$ denotes a diagonal matrix with the elements of $S$ on the diagonal and the dot denotes differentiation with respect to time $t$, see for instance Arino et al in \cite{arino2005multi}. The parameter $\beta$ is the disease transmission rate and $\gamma$ is the recovery rate. The SEIR model in (\ref{Eq:SEIR-model}) is scaled to fractions of the total initial population $N_p$ such that
\begin{eqnarray}\label{Eq:SEIR-modelNorm}
    S + E + I + R &=& \nu \\
    \sum_j^n \nu_j &=& 1\ ,
\end{eqnarray}
where $\nu \in (0,1)^n$ is the fraction of the total population in group $j$.
The population of susceptible at time $t$ equals $S_N(t)=N_p S(t)$, similarly for the exposed $E_N(t)=N_p E(t)$, the infectious $I_N(t)=N_p I(t)$, and for the recovered $R_N(t)=N_p R(t)$. Subscript $N$ in $S_N, E_N, I_N$, and $R_N$ refers to the actual number of susceptible, exposed, infected and recovered. We will later look at special cases to simplify certain calculations.

The contact matrix $\beta\in(0,\infty)^{n\times n}$ holds the rate of interactions between different groups. Such stratification could be ages (interactions between young and old), physical location (different cities or countries), or species (such as mosquitoes and humans) \cite{SIR2, SIR4, SIR5}. To study the effects of adaptive behaviour, we promote the contact matrix $\beta$ to be a function of time and infection numbers in the following way:
\begin{eqnarray}\label{Eq:betaInt}
	\beta_{jk}(t) &=& (\beta_0)_{jk} - \sum_{m=1}^{n}\int\limits_{-\infty}^{t} \alpha_{jkm}(t-\tau) I_m(\tau) d\tau\ .
\end{eqnarray}
That is, an integration kernel consisting of a linear series of the functions $\alpha_k$, where $\alpha_k$ is proportional to a product of $u^{(k-1)} e^{-\sigma u}$ with $u=t-\tau$ and an oscillating harmonic part.
The first term is considered independent of time, and the second term is responsible for memory effects in the dynamics such as adaptive behavior. The 3-tensor kernel $\alpha$ may in general be very complicated as long as it satisfies
\begin{eqnarray}
	\int\limits_{0}^{\infty}\vert(\alpha)_{jkm}(t)\vert dt < \infty,\quad m=1,2,\cdots,M .
\end{eqnarray}
The kernel will typically decrease as $u$ increases and otherwise the overall shape of $\beta$ can be chosen to best represent the given data, e.g., whether it goes to 0 at the origin, which determines whether the feedback is immediate. In mathematical terms, we have a lot of freedom to choose a family of functions for the kernel.

We shall consider a kernel with an oscillating part for modelling e.g. seasonal variations or weekly variations due to shifts in behaviour between work and leisure time in the weekends. We use a kernel previously investigated by Ponosov et al. \cite{ponosov2004thew} but now adding an oscillating term as follows
\begin{eqnarray}
\label{Eq:kernel-complex}
    \alpha_{k}(u) &=& c_{k} \frac{\sigma}{(k-1)!}u^{k-1}e^{-\sigma u}\big[\frac{1}{2} + (\varepsilon_k + i\mu_k) e^{i \omega u} \big] + \text{cc}
 \quad \mathrm{where} \quad u=t-s \; .
\end{eqnarray}
The $cc$ stands for complex conjugation of the preceding terms. We have here suppressed the matrix indices for readability, but all parameters can easily be given more indices if needed. We will also focus on a non-stratified model in this paper.

Note that we can consider a kernel with explicit time dependence $\alpha_{T}(t, t-\tau)$ as long as it is of the form $\alpha_T(t, t-\tau) = c(t) \tilde{\alpha}_T(t-\tau)$. This will allow $c(t)$ to be pulled outside the integral and handled as a part of the rest of the differential equations. This is useful, either when looking at seasonal changes \cite{EbolaSubExp1, EbolaSubExp2, EbolaSubExp3} or sub-exponential growth \cite{SubExp}. It makes sense to do this in conjunction with a time-dependent $\beta_0$. To include oscillations in the kernel, depending on the delay time u=t-s, may seem less obvious. But the oscillations in the kernel depending on $u$ could arise from people, who adjust their behavior today on their experience same time last year or same day last week. An example could be risky behavior due to gathering last weekend and in the current weekend people wish to behave less risky by staying home. From a mathematical point of view it is of interest that the linear chain trick can be extended to the case of an oscillating kernel as in Eq. (\ref{Eq:kernel-complex}). Furthermore, such oscillating kernels appear in physics and here we can mention the delayed Raman response in nonlinear optical fibers \cite{dudley2010supercontinuum}.

The parameters $\sigma \in (\mathbb{R}_+)^{n\times n}$ are positive real numbers, and the positive integers $k$ takes the values $k=1, 2, ... \, N$ with $N \in \mathbb{N}_+$. Furthermore, $c_k, \omega, \varepsilon_k, \mu_k \in \mathbb{R}^{n\times n}$ and $n \in \mathbb{N}$.

\subsection{Rewriting the Integral Kernel as a Set of ODEs}
In order to solve Equation (\ref{Eq:SEIR-model}) numerically using Equation (\ref{Eq:kernel-complex}) we apply the linear chain trick transforming the integro-differential equation into a set of ordinary differential equations. This transformation implies two advantages. First, the system of integro-differential equations in (\ref{Eq:SEIR-model}) can be solved numerically using ordinary differential equation solvers without invoking numerical methods for finding the integral parts. Secondly, stability analysis of equilibrium points for (\ref{Eq:SEIR-model}) can be conducted using methods from ordinary differential equations. We start by writing
\begin{equation}
    \label{Eq:alphak-complex}
    \alpha_{k}(u) = G_0^k(u) + G_1^k(u) + G_2^k(u) \; ,
\end{equation}
\noindent where we have introduced
\begin{equation}
\label{Eq:Gnk}
 \left.\begin{aligned}   
   & G_0^k(u) = \frac{c_k \sigma}{(k-1)!}u^{k-1}e^{-\sigma u} \\  
   & G_1^k(u) = \frac{c_k (\varepsilon_k + i\mu_k) \sigma}{(k-1)!}u^{k-1} e^{\left(-\sigma + i\omega\right) u} \\
   & G_2^k = \left(\overline{G_1^k}\right) = \frac{c_k (\varepsilon_k - i\mu_k) \sigma}{(k-1)!}u^{k-1} e^{\left(-\sigma - i\omega\right) u}
\end{aligned} \right.
\end{equation}

\noindent From the Equation (\ref{Eq:betaInt}) we observe that we need to calculate integrals of the form
\begin{equation}
    \label{Eq:IntG}
    z_k(t) = \int\limits_{-\infty}^{t} \alpha_k(t-s) I(s) ds  = z_{k}^{(0)}(t) + z_{k}^{(1)}(t) + z_{k}^{(2)}(t) \; ,
\end{equation}

\noindent where

\begin{equation}
    \label{Eq:zkj}
    z_{k}^{(0)}(t) = \int\limits_{-\infty}^{t} G_0^k(t-s) I(s) ds \; , \quad z_{k}^{(1)}(t) = \int\limits_{-\infty}^{t} G_1^k(t-s) I(s) ds \; , \quad z_{k}^{(2)}(t) = \overline{z_{k}^{(2)}}(t)\; ,
\end{equation}
and the bar indicates the complex conjugate.

\noindent Applying the linear chain trick as presented in \cite{ponosov2004thew} we can find differential equations for $z_{k}^{(j)}(t)$, $j=0,1,2$, by differentiating the integrals in (\ref{Eq:zkj}). The integrals in the delay differential equation (\ref{Eq:betaInt}) can thus be replaced by a set of differential equations for $z_{k}^{(j)}(t)$, using the particular form for $\alpha_k$ in (\ref{Eq:kernel-complex}). Differentiating $z_{k}^{(0)}(t)$ we get

\begin{equation}
    \label{Eq:dzk1-dt}
    \frac{dz_{k}^{(0)}(t)}{dt} = G_0^k(0) I(t) + \int\limits_{-\infty}^{t} \frac{d G_0^k(t-s)}{dt} I(s) ds \; .
\end{equation}

\noindent For $k=1$ we have from Equation (\ref{Eq:Gnk}) that $G_0^1(u) = c_1 \sigma e^{-\sigma u}$ and accordingly $G_0^1(0) = c_1 \sigma$ and $G_0^1(u) \rightarrow 0 $ for $u \rightarrow \infty$. For $k=2, 3, ... , N$ we have that $G_0^k(0) = 0$ and $G_0^k(u) \rightarrow 0 $ for $u \rightarrow \infty$. We can easily differentiate  $G_0^k$ in (\ref{Eq:Gnk}) with respect to $t$ and use the definition of $z_{k}^{(0)}(t)$ in (\ref{Eq:zkj}) to obtain the differential equations for $z_{1}^{(0)}(t)$ and $z_{k}^{(0)}(t)$, for $k=2, 3, ... , N$

\begin{equation}
\label{Eq:dzk1a}
 \left.\begin{aligned}
   &  \frac{d z_{1}^{(0)}(t) }{dt} = c_1 \sigma I(t) - \sigma z_{1}^{(0)}(t)\\  
   &  \frac{d z_{k}^{(0)}(t) }{dt} = \frac{c_{k}}{c_{k-1}} z_{k-1}^{(0)}(t) - \sigma z_{k}^{(0)}(t) \quad \mathrm{for} \quad k=2,3, ... , N \; .
\end{aligned} \right.
\end{equation}

\noindent We now continue by deriving a differential equation for $z_{k}^{(1)}(t)$ following the procedure for $z_{k}^{(0)}(t)$. We have

\begin{equation}
    \label{Eq:dzk2-dt}
    \frac{d z_{k}^{(1)}(t) }{dt} = G_1^k(0) I(t) + \int\limits_{-\infty}^{t} \frac{d G_1^k(t-s)}{dt} I(s) ds \; .
\end{equation}

\noindent From Equation (\ref{Eq:Gnk}) using $k=1$ we have $G_1^1(u) = \sigma c_1 (\varepsilon_1 + i\mu_1) e^{\left(-\sigma + i\omega\right) u}$ and thus $G_1^1(0) = \sigma c_1 (\varepsilon_1 + i\mu_1)$ and $G_1^1(u) \rightarrow 0 $ for $u \rightarrow \infty$. Furthermore, for $k=2, 3, ... , N$ we have $G_1^k(0) = 0$ and $G_1^k(u) \rightarrow 0 $ for $u \rightarrow \infty$. We differentiate $G_1^k$ in (\ref{Eq:Gnk}) with respect to $t$ and use the definition of $z_{k}^{(1)}(t)$ in (\ref{Eq:zkj}) to obtain the differential equations governing $z_{k}^{(1)}(t)$, for $k=1, 2, 3, ... , N$

\begin{equation}
\label{Eq:dzk2a}
 \left.\begin{aligned}   
   &  \frac{d z_{1}^{(1)}(t) }{dt} = \sigma c_1 (\varepsilon_1 + i\mu_1) I(t) + \left(-\sigma + i\omega\right) z_{1}^{(1)}(t)\\  
   &  \frac{d z_{k}^{(1)}(t) }{dt} = \sigma \frac{c_k(\varepsilon_k + i\mu_k)}{c_{k-1}(\varepsilon_{k-1} + i\mu_{k-1})} z_{k-1}^{(1)}(t) + \left(-\sigma + i\omega\right) z_{k}^{(1)}(t) \quad \mathrm{for} \quad k=2,3, ... , N \; .
\end{aligned} \right.
\end{equation}

\noindent Differential equations for $z_{k}^{(2)}(t)$, $k=1,2, \dots , N$, are easily obtained from noting that $z_{k}^{(2)}(t) = \overline{z_{k}^{(1)}}(t)$ and accordingly we obtain

\begin{equation}
\label{Eq:dzk3a}
 \left.\begin{aligned}   
   &  \frac{d z_{1}^{(2)}(t) }{dt} = \sigma c_1 (\varepsilon_1 - i\mu_1) I(t) + \left(-\sigma - i\omega\right) z_{13}(t)\\  
   &  \frac{d z_{k}^{(2)}(t) }{dt} = \sigma z_{k-1}^{(2)}(t) + \left(-\sigma - i\omega\right) z_{k}^{(2)}(t) \quad \mathrm{for} \quad k=2,3, ... , N \; .
\end{aligned} \right.
\end{equation}

\noindent Collecting the above, our aim is to solve the system of differential equations in (\ref{Eq:dzk1a}), (\ref{Eq:dzk2a}) and (\ref{Eq:dzk3a}). Initial conditions are specified for $S$, $I$ and $R$. The initial conditions for $z_{kj}$, $k=1,2, ...,N$, and $j=1,2,3$, must also be specified. We have

\begin{equation}\label{Eq:SIR-delay-IC}
	\left.\begin{aligned}
		z_{k}^{(0)}(0) \ =&\ \int\limits_{-\infty}^{0} G_0^k(-s) I(s) ds \\
		z_{k}^{(1)}(0) \ =&\ \int\limits_{-\infty}^{0} G_{1}^k(-s) I(s) ds \\
		z_{k}^{(2)}(0) \ =&\ \int\limits_{-\infty}^{0} G_{2}^k(-s) I(s) ds
	\end{aligned}\right.
\end{equation}

\noindent In these expressions we need to specify $I(s)$ for $s \in ( -\infty ; 0]$, which is typically very difficult in realistic systems. Though this is of course also a problem if one wants to determine the initial conditions for $I$ in an SEIR-model without adaptive behaviour. In both cases, fitting infection data to a polynomial is tenable \cite{EkspertRapporterDK}. In our simulations, we assume that $I(s)<0$ for $s \in ( -\infty ; 0]$, simplifying this problem.

\subsection{Numerics}\label{Sec:Numerics}

\noindent For illustration, we restrict ourselves to the simpler form of the kernel
\begin{eqnarray}\label{Eq:Kernel}
	\alpha(u) &=& \frac{c_0}{2} u e^{-\sigma t}\left(1 + (\epsilon + i\mu) e^{i \omega u} \right) + \text{cc}\ .
\end{eqnarray}
That is, Equation (\ref{Eq:Gnk}) for $k=2$. For this specific kernel \eqref{Eq:Kernel}, the linear chain trick has the following form.
We express it in the following real functions
\begin{equation}\label{Eq:AuxVar}
	\left.\begin{aligned}
	\beta^0_R(t) = \frac{c_0}{2} e^{(-\sigma + i \omega) t} + \text{cc} & \quad, \quad \beta^0_I(t) = i\frac{c_0}{2} e^{(-\sigma + i \omega) t} + \text{cc}\ ,\\
    \beta^1_R(t) = \frac{c_0}{2} t e^{(-\sigma + i \omega) t} + \text{cc} & \quad, \quad \beta^1_I(t) = i \frac{c_0}{2} t e^{(-\sigma + i \omega) t} + \text{cc}\ ,\\
	u^{(0)}(t) = \sum\limits_{m=1}^{M} \int\limits_{-\infty}^{t} \beta^0_R(t-\tau) I_m(\tau) d\tau &\quad , \quad v^{(0)}(t) = \sum\limits_{m=1}^{M} \int\limits_{-\infty}^{t} \beta^0_I(t-\tau) I_m(\tau) d\tau\ ,\\
	u^{(1)}(t) = \sum\limits_{m=1}^{M} \int\limits_{-\infty}^{t} \beta^1_R(t-\tau) I_m(\tau) d\tau& \quad ,\quad v^{(1)}(t) = \sum\limits_{m=1}^{M} \int\limits_{-\infty}^{t} \beta^1_I(t-\tau) I_m(\tau) d\tau\ ,\\
	\beta_H(t) = c_0 e^{-\sigma t}&\quad, \quad \widehat{\beta}(t) = \sum\limits_{m=1}^{M} \int\limits_{-\infty}^{t} \beta_H(t-\tau) I_m(\tau) d\tau\ .
	\end{aligned}\right.
\end{equation}
We will later set $n=1$, but as some results are obtainable for general $n$, we keep it for now.
The following differential equation for $\beta$ in Equation \eqref{Eq:betaInt} is
\begin{eqnarray}
	\dot{\beta}_{jk} &=& -\sum_{m=1}^n \left(\overbrace{\alpha(0)}^{=0}I_m(t) + \int\limits_{-\infty}^{t} \partial_t \alpha(t-\tau) I_m(\tau) d\tau \right) + \text{cc}\nn\\
	&=& -\frac{c_0}{2} \sum_{m=1}^n \int\limits_{-\infty}^{t} e^{-\sigma (t-\tau)}\left[\left(1 + (\epsilon + i\mu) e^{i \omega(t-\tau)} \right)  \left( 1 -\sigma t \right) + i\omega t (\epsilon+i\mu) e^{i\omega (t-\tau)} \right]I_m(\tau) d\tau + \text{cc}\nn\\
	&=& -\sigma \left(\beta_{jk} - (\beta_0)_{jk} \right) - \widehat{\beta} - \omega (\epsilon - \mu)u^{(1)} - \omega (\epsilon + \mu)v^{(1)}
\end{eqnarray}
	and similarly for the helper functions we get
\begin{equation}
	\left.\begin{aligned}
	\dot{\widehat{\beta}} &=& c_0 \left(1+\epsilon\right) \sum\limits_{m=1}^{n}I_m - \sigma \widehat{\beta}\\
	\dot{u}^{(1)} &=& u^{(0)} - \sigma u^{(1)} + \omega v^{(1)}\\
	\dot{v}^{(1)} &=& v^{(0)} - \sigma v^{(1)} - \omega u^{(1)}\\
	\dot{u}^{(0)} &=&  c_0 \left(1+\epsilon\right) \sum\limits_{m=1}^{n}I_m - \sigma u^{(0)} + \omega v^{(0)}\\
	\dot{v}^{(0)} &=& c_0 \left(1+\epsilon\right) \sum\limits_{m=1}^{n}I_m - \sigma v^{(0)} - \omega u^{(0)}
	\end{aligned}\right.
\end{equation}
This gives us a set of differential equations that describes our system
\begin{equation}\label{Eq:SEIR-model-Full}
	\left.\begin{aligned}
		\dot{S}\ =&\ - \diag(S)\beta I\\
		\dot{E}\ =&\   \diag(S) \beta I - \eta E\\
		\dot{I}\ =&\ \eta E - \gamma I\\
		\dot{R}\ =&\ \gamma I\\
		\dot{\beta}\ =&\ -\sigma \left(\beta - (\beta_0) \right) - \widehat{\beta} - \omega (\epsilon - \mu)u^{(1)} - \omega (\epsilon + \mu)v^{(1)}\\
		\dot{\widehat{\beta}}\ =&\ c_0 \left(1+\epsilon\right) \sum\limits_{m=1}^{n}I_m - \sigma \widehat{\beta}\\
		\dot{u}^{(1)}\ =&\ u^{(0)} - \sigma u^{(1)} + \omega v^{(1)}\\
		\dot{v}^{(1)}\ =&\ v^{(0)} - \sigma v^{(1)} - \omega u^{(1)}\\
		\dot{u}^{(0)}\ =&\ c_0 \left(1+\epsilon\right)\sum_{m=1}^{n} I_m - \sigma u^{(0)} + \omega v^{(0)}\\
		\dot{v}^{(0)}\ =&\ c_0 \left(1+\epsilon\right) \sum_{m=1}^{n} I_m - \sigma v^{(0)} - \omega u^{(0)}
	\end{aligned}\right.
\end{equation}
with the initial conditions $\beta = \beta_0$ and $\widehat{\beta} = u^{(1)} = v^{(1)} = u^{(0)} = v^{(0)} = 0$. That is, we assume there has been no infection before $t=0$.

\begin{figure}[t]
	\centering
	\includegraphics[width=\linewidth]{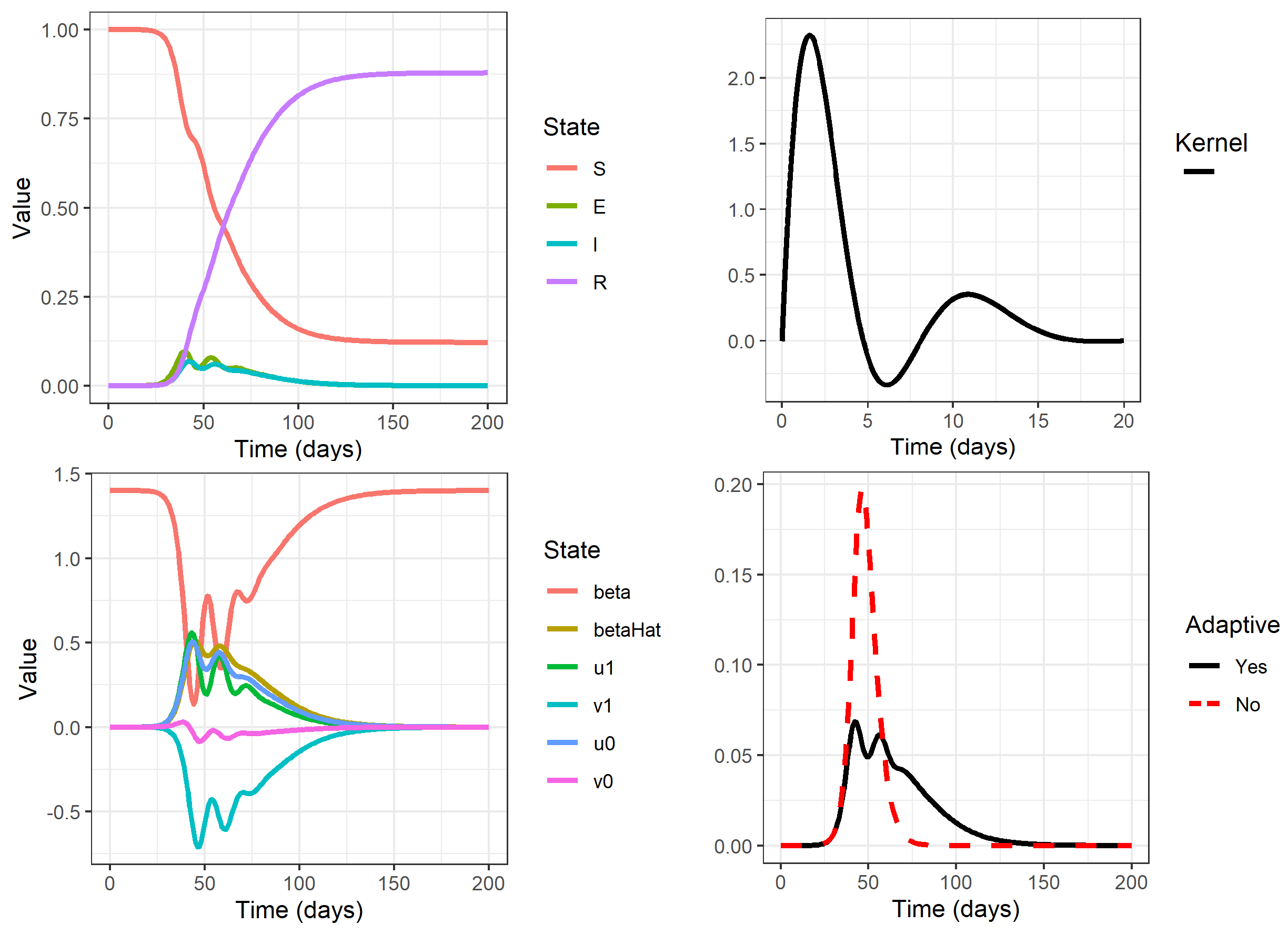}
	\caption{Plots of SEIR-model with adaptive behaviour. Parameter values are taken from Table \ref{Tab:ExampleParameters} and using $n=1$, that is we consider $S$, $E$, $I$ and $R$ to be scalars. \textbf{Top left:} The classical variables $S$, $E$, $I$, and $R$. \textbf{Bottom left:} The auxiliary variables from Equation \eqref{Eq:AuxVar} that pertain to the contact rate. Note that $\beta$ stays positive for all times. \textbf{Top right:} Plot of the integral kernel from Equation \eqref{Eq:Kernel} used in the simulation. \textbf{Bottom right:} Comparison of the $I$-state for an SEIR-model with the same time parameters, but with and without adaptive behaviour. (The one without adaptive behaviour simply has $c_0=0$.}
	\label{Fig:SEIRdelayCompare}
\end{figure}

Numerical results illustrating the effect of the delay are shown in Figure \ref{Fig:SEIRdelayCompare}, using the parameter values in Table \ref{Tab:ExampleParameters}. We consider the scalar case of the dependent variables $S$, $E$, $I$ and $R$ corresponding to $n=1$.

\begin{table}[h]
    \centering
    \begin{tabular}{llll}
    \toprule
       $\eta = 0.25 \: \mathrm{d}^{-1}$ & \quad $c_0 = 1.3 \: \mathrm{d}^{-1}$  &  \quad $\lambda_R = 0.4 \: \mathrm{d}^{-1}$ & \quad $\varepsilon = 1.5$\\
       $\gamma = 0.3 \: \mathrm{d}^{-1}$ & \quad $\beta_0 = 1.4 \: \mathrm{d}^{-1}$ & \quad $\lambda_I = 0.5 \: \mathrm{d}^{-1}$  & \quad $\mu  = -0.1$ \\
    \bottomrule
    \end{tabular}
    \caption{Parameter values used in Figure \ref{Fig:SEIRdelayCompare}. The time unit is one day denoted d. These parameters are chosen to be illustrative rather than realistic.}
    \label{Tab:ExampleParameters}
\end{table}

\subsection{Equilibrium Points and Stability}
The results for $t\to \infty$ are of course still determined by depletion of susceptibles, but as long as the number of susceptibles is assumed to be roughly constant, the memory effects have interesting consequences, such as stability of the number of infections on short time scales. This explains why a contact number around 1 is observed more often in real-world systems than a traditional exponential model would suggest. We investigate these in detail with the example kernel in Equation \eqref{Eq:Kernel}.\\
\newline
Leaving out the recovered state $R$ through normalization and assuming approximately constant number of susceptibles $S$, the Jacobian is
\begin{eqnarray}\label{Eq:JacobianFull}
	J &=& \left(\begin{array}{c|cccccccc}
		& \partial_{E_j} & \partial_{I_j} & \partial_{\beta_{jm}} & \partial_{\widehat{\beta}} & \partial_{u^{(1)}} & \partial_{v^{(1)}} & \partial_{u^{(0)}} & \partial_{v^{(0)}}\\
		\hline
		\dot{E}_{k} & -\eta & S_k \beta_{kj} & S_j I_m & 0 & 0 & 0 & 0 & 0\\
		\dot{I}_{k} & \eta & -\gamma & 0 & 0 & 0 & 0 & 0 & 0\\
		\dot{\beta}_{kn} & 0 & 0 & -\sigma \delta_{jk} \delta_{mn} & -1 & -\omega (\epsilon - \mu) & -\omega (\epsilon + \mu) & 0 & 0\\
		\dot{\widehat{\beta}} & 0 & c_0 \left(1+\epsilon\right) & 0 & -\sigma & 0 & 0 & 0 & 0\\
		\dot{u}^{(1)} & 0& 0 & 0 & 0 & -\sigma & \omega & 1 & 0\\
		\dot{v}^{(1)} & 0 & 0 & 0 & 0 & -\omega & -\sigma & 0 & 1\\
		\dot{u}^{(0)} & 0& c_0 \left(1+\epsilon\right) & 0 & 0 & 0 & 0 & -\sigma & \omega\\
		\dot{v}^{(0)} & 0 & c_0 \left(1+\epsilon\right) & 0 & 0 & 0 & 0 & -\omega & -\sigma
	\end{array}\right)
\end{eqnarray}
Note that $u$, $v$, and $\widehat{\beta}$ do not need indices, but $\beta$ and $\beta_0$ do because we may want $n>1$.

\subsubsection{Disease Free}
\noindent Let us start with the trivial disease-free equilibrium point
\begin{equation}
	\left.\begin{aligned}
		E^*_j = I^*_j = \widehat{\beta}^*_{jk}\ =\ (u^{(1)})^* = (v^{(1)})^*\ =&\ (u^{(0)})^* = (v^{(0)})^* = 0\\
		\beta^*\ =&\ \beta_0
	\end{aligned}\right.
\end{equation}
The Jacobian \eqref{Eq:JacobianFull} reduces to
\begin{eqnarray}
	J_{DF} &=& \left(\begin{array}{cccccccc}
		-\eta & S_k (\beta_0)_{kj} & 0 & 0 & 0 & 0 & 0 & 0\\
		\eta & -\gamma & 0 & 0 & 0 & 0 & 0 & 0\\
		0 & 0 & -\sigma \delta_{jk} \delta_{mn} & -1 & -\omega (\epsilon - \mu) & -\omega (\epsilon + \mu) & 0 & 0\\
		0 & c_0 \left(1+\epsilon\right) & 0 & -\sigma & 0 & 0 & 0 & 0\\
		0 & 0 & 0 & 0 & -\sigma & \omega & 1 & 0\\
		0 & 0 & 0 & 0 & -\omega & -\sigma & 0 & 1\\
		0 & c_0 \left(1+\epsilon\right) & 0 & 0 & 0 & 0 & -\sigma & \omega\\
		0 & c_0 \left(1+\epsilon\right) & 0 & 0 & 0 & 0 & -\omega & -\sigma
	\end{array}\right)
\end{eqnarray}
Note that all blocks but $S_j (\beta_0)_{jk}$ are proportional to the identity matrix, so we may diagonalize that block on its own and relate the eigenvalues $x_{S\beta_0}$ of $\diag(S)\beta_0$ to those of $J_{DF}$. The largest eigenvalue (i.e.\ the one that potentially can be positive) is
\begin{eqnarray}
	EV_{\rm max} &=& \frac{-\eta - \gamma + \sqrt{\eta^2 - 2 \eta \gamma + \gamma^2 + 4\eta x_{S\beta_0}}}{2}
\end{eqnarray}
So the condition is $x_{S\beta_0} < \gamma$ for the disease-free equilibrium point to be stable. This corresponds to a reproduction number

\begin{equation}
    \label{Rnd}
    R_{ND} = \frac{S\beta_0}{\gamma}
\end{equation}

\noindent above 1 in a normal SEIR-model. This is significant, because it shows that feedback of the kind \eqref{Eq:Kernel} cannot change whether or not there will be an outbreak, only the severity of it.

\subsubsection{Equilibrium Point During Outbreak}
\noindent It turns out that there is also an equilibrium point during an outbreak
\begin{equation}
	\left.\begin{aligned}
		\widehat{\beta}^*\ = &\ \frac{c_0 \left(1+\epsilon\right)}{\sigma} \sum_{m=1}^{M} I^*_m\\
		(u^{(0)})^*\ = &\ \frac{c_0\left(\sigma + \omega\right) \left(1+\epsilon\right)}{{\sigma}^2 + {\omega}^2} \left(\sum_{m=1}^{M} I^*_m\right)\\
		(v^{(0)})^*\ = &\ \frac{c_0\left(\sigma - \omega\right) \left(1+\epsilon\right)}{{\sigma}^2 + {\omega}^2} \left(\sum_{m=1}^{M} I^*_m\right)\\
		(u^{(1)})^*\ =&\ \frac{\sigma (u^{(0)})^* + \omega (v^{(0)})^*}{{\sigma}^2 + {\omega}^2}\ =\  \frac{c_0 \left(1+\epsilon\right) \left({\sigma}^2 - {\omega}^2 + 2 \sigma \omega\right)}{\left({\sigma}^2 + {\omega}^2\right)^2} \left(\sum_{m=1}^{M} I^*_m\right)\\
		(v^{(1)})^*\ =&\ \frac{\sigma (v^{(0)})^* - \omega (u^{(0)})^*}{{\sigma}^2 + {\omega}^2}\ =\ \frac{c_0 \left(1+\epsilon\right) \left({\sigma}^2 - {\omega}^2 - 2 \sigma \omega\right) }{\left({\sigma}^2 + {\omega}^2\right)^2} \left(\sum_{m=1}^{M} I^*_m\right) \\
		\beta^*\ =&\ \beta_0 - \frac{\widehat{\beta}^* + \omega (\epsilon - \mu)(u^{(1)}) + \omega (\epsilon + \mu)(v^{(1)}) }{\sigma}\\
		E^*\ =&\ \frac{\gamma}{\eta}I^*\\
		\beta_0 I^*\ =&\ \diag\left(S\right)\inv \gamma I^* + c_0 \left(1+\epsilon\right) \left(\frac{1}{{\sigma}^2} + 2 \frac{\omega}{\sigma} \frac{\epsilon \left({\sigma}^2 - {\omega}^2\right) - 2\sigma\omega\mu}{\left({\sigma}^2 + {\omega}^2\right)^2}\right) \left(\sum_{m=1}^{M} I^*_m\right)I^*
	\end{aligned}\right.
\end{equation}
The last condition on $I$ is very difficult to solve in general, especially because the quantity $\sum\limits_{m=}^{n}I^*_m$ is not invariant under diagonalization of $\beta_0$. We therefore continue with a single group (i.e., $n=1$) to see what properties can be divined in this case. This reduces the equilibrium point equations to
\begin{equation}\label{Eq:FixedPoint}
	\left.\begin{aligned}
	    I^* \ =&\ \frac{\beta_0 - \gamma/S}{c_0 \left(1+\epsilon\right) \left(\frac{1}{{\sigma}^2} + 2 \frac{\omega}{\sigma} \frac{\epsilon \left({\sigma}^2 - {\omega}^2\right) - 2\sigma\omega\mu}{\left({\sigma}^2 + {\omega}^2\right)^2}\right)}\\
		\widehat{\beta}^*\ = &\ \frac{c_0 \left(1+\epsilon\right)}{\sigma} I^*\\
		E^*\ =&\ \frac{\gamma}{\eta}I^*\\
		(u^{(0)})^*\ = &\ \frac{c_0 \left(1+\epsilon\right)\left(\sigma + \omega\right)}{{\sigma}^2 + {\omega}^2} I^*\\
		(v^{(0)})^*\ = &\ \frac{c_0 \left(1+\epsilon\right)\left(\sigma - \omega\right)}{{\sigma}^2 + {\omega}^2} I^*\\
		(u^{(1)})^*\ =&\ \frac{c_0\left(1+\epsilon\right)\left({\sigma}^2 - {\omega}^2 + 2 \sigma \omega\right) }{\left({\sigma}^2 + {\omega}^2\right)^2} I^*\\
		(v^{(1)})^*\ =&\ \frac{c_0\left(1+\epsilon\right)\left({\sigma}^2 - {\omega}^2 - 2 \sigma \omega\right) }{\left({\sigma}^2 + {\omega}^2\right)^2} I^* \\
		\beta^*\ =&\ \frac{\gamma}{S}
	\end{aligned}\right.
\end{equation}
We require $I>0$, which means that the non-damped reproduction number has to be larger than 1, corresponding to an ongoing epidemic. (By non-damped we mean $c_0 = 0$ where there is no feedback.)
It is clear that below this point, we transition to the disease-free equilibrium point.
This assumes that the denominator in $I^*$ is positive. Note that negative denominator does not make physical sense, as we then get more infected when the contact rate is lowered.

We also implicitly assume that $I^*\ll S$. Otherwise the change in susceptibles will play a role in the dynamics.\\
\newline
For the sake of stability analysis, we start by looking at a simplified case where $\omega=0$, and then treat the full version numerically. First we take the case $\omega = 0$. As discussed above, this is a very physiologically relevant case. For simplicity, we also set $\epsilon=\mu=0$ as these may otherwise simply be absorbed in $c_0$.
Here the equilibrium point for $M=1$ is
\begin{equation}
	\left.\begin{aligned}
		I^*\ =&\ \frac{{\sigma}^2}{c_0 S}\left(R_{ND}-1\right) \gamma\\		
  E^*\ =&\ \frac{\gamma}{\eta}I^*\\
		\beta^*\ =&\ \frac{\gamma}{S}\\
		\widehat{\beta}^*\ =\ (u^{(0)})^*\ =\ (v^{(0)})^*\ = &\ \frac{\sigma}{S}\left(R_{ND} - 1 \right)  \gamma\\
		(u^{(1)})^*\ =&\ (u^{(0)})^*/\sigma\\
		(v^{(1)})^*\ =&\ (v^{(0)})^*/\sigma
	\end{aligned}\right.
\end{equation}
Note an important, but perhaps unsurprising feature here. As $c_0\sim\sigma^2$ for a normalized version of the kernel in \eqref{Eq:Kernel}, the stable level of infection $I$ depends only on the integral of the kernel. So epidemic non-pharmaceutical interventions may be spread out over a period of time, or they may be strict and fast. It gives the same level of infection.

The variables $u^{(0)}$, $v^{(0)}$, $u^{(1)}$, and $v^{(1)}$ drop out of the dynamics, and the relevant part of the Jacobian \eqref{Eq:JacobianFull} therefore reduces to
\begin{eqnarray}
	J_{\omega=0} &=& \left(\begin{array}{cccc}
		-\eta & \gamma & \frac{{\sigma}^2}{c_0} \left(R_{ND} - 1 \right) \gamma & 0\\
		\eta & -\gamma & 0 & 0 \\
		0 & 0 & -\sigma & -1 \\
		0 & c_0 & 0 & -\sigma 
	\end{array}\right)\ .
\end{eqnarray}
We will use the Routh-Hurwitz theorem to determine the stability of the system, specifically the formulation from \cite{RouthHurwitz}. The characteristic polynomial for the Jacobian is
\begin{eqnarray}
	P^J_{\omega = 0}(x) &=& x^4 + a_1 x^3 + a_2 x^2 + a_3 x  + a_4\ .
\end{eqnarray}
where
\begin{equation}
	\left.\begin{aligned}
		a_1\ =&\ \eta + \gamma + 2 \sigma\\
		a_2\ =&\ 2 \sigma (\eta + \gamma + \sigma)\\
		a_3\ =&\ \sigma^2(\eta + \gamma)\\
		a_4\ =&\ \eta \sigma^2 \left(R_{ND} - 1 \right) \gamma
	\end{aligned}\right.
\end{equation}
These are all positive, because $R_{ND} = \beta_0 S / \gamma\geq 1$. Using the definitions from \cite{RouthHurwitz}
\begin{equation}\label{Eq:RH_det_nonOsc}
	\begin{aligned}
		\mathbf{D_1} = a_1\ ,\ 
		\mathbf{D_2} = \left(\begin{matrix}
			a_1 & a_3\\
			1 & a_2
		\end{matrix}\right)\ ,\ 
		\mathbf{D_3} &= \left(\begin{matrix}
			a_1 & a_3 & 0\\
			1 & a_2 & a_4\\
			0 & a_1 & a_3
		\end{matrix}\right)\ ,\
		\mathbf{D_4} = \left(\begin{matrix}
			a_1 & a_3 & 0 & 0\\
			1 & a_2 & a_4 & 0\\
			0 & a_1 & a_3 & 0\\
			0 & 1 & a_2 & a_4
		\end{matrix}\right) .
	\end{aligned}
\end{equation}
From the condition $\det(\mathbf{D_4})>0$ we find
\begin{eqnarray}\label{Eq:R_ND-cond}
	&& 1 < \frac{S(0) \beta_0}{\gamma} = R_{ND} < \nn \\ 
 && \frac{\eta \gamma (\eta + \gamma)^2 + 2 (\eta + \gamma) (\eta^2 + 4 \eta \gamma + \gamma^2) \sigma + 4 (\eta^2 + 3 \eta \gamma + \gamma^2) \sigma^2 + 2 (\eta + \gamma) \sigma^3}{\eta \gamma (\eta + \gamma + 2 \sigma)^2}\ .
\end{eqnarray}
There is also a root of $\det(\mathbf{D_3})$ at this point, but as $\sign(a_1) = \sign(a_3)$, there is still a Hopf bifurcation when this criterion is not satisfied \cite{RouthHurwitz}.

\begin{figure}
	\centering
	\hspace{-50pt}\includegraphics[width=0.9\linewidth]{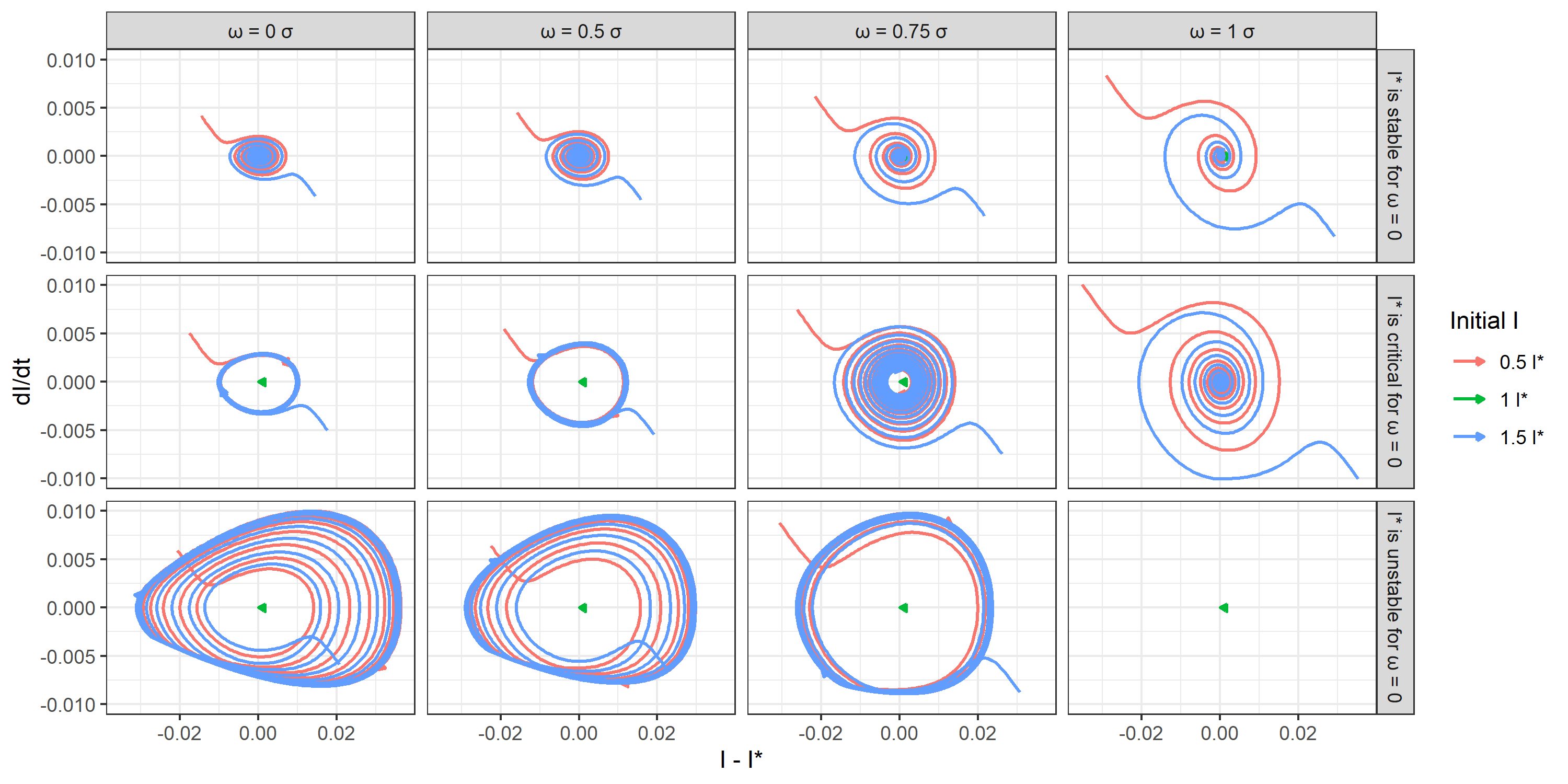}
	\caption{Phase space and stability for $\omega$ and $\beta_0$ using $n=1$. We keep $S$ constant to illustrate the local stability in time. We choose three set of parameters that are unstable, critical, and unstable for $\omega=0$, using Equation (\ref{Eq:R_ND-cond}), and then vary $\omega$ to analyse changes in stability.
	The parameters are $\gamma = \eta = 1/3.5,\ c_0 = 5.5,\ \sigma = 0.5,\ S = 0.8,\ \epsilon=\mu=0.5$ for all the configurations and $\beta_0 = 1.319,\ 1.519,\ 1.719$ for the stable, critical, and unstable configurations respectively (top, middle, and bottom row respectively), corresponding to non-damped generation numbers of 3.70, 4.25, and 4.81 respectively.
	The initial conditions are chosen according to the equilibrium point from Equation (\ref{Eq:FixedPoint}), but with $I$ varied to show convergence of different paths. We have omitted runs that exit the interval $E+I\in[0,1]$ as they are unphysical. These typically display large oscillations. We also use $\max(\beta,0)$ instead of $\beta$ in the RHS of Equation \ref{Eq:SEIR-model-Full}.
	We see that $\omega$ does indeed allow changes in stability, but mostly around the critical point.}
	\label{Fig:AllStability}
\end{figure}

Note an interesting property of Equation (\ref{Eq:R_ND-cond}). The stability of the equilibrium point is independent of $c_0$, and the interpretation seems to be the following: As long as the feedback is non-zero, it cannot change the stability of the equilibrium point, only the position. A sign change would give rise to an equilibrium point at negative $I$, which in turn would change the sign of the integral in Equation \ref{Eq:betaInt}. This is of course purely formal as $I<0$ is unphysical. It is merely to explain the absence of $c_0$ in the stability condition.
Numerics shows that the transition is between a stable equilibrium point and a stable limit cycle, see Figure \ref{Fig:AllStability}, left column.

\subsubsection{Numerical Stability Analysis of Full Model}
We start by choosing parameters just above, below, and at the critical point in Equation (\ref{Eq:R_ND-cond}) for $\omega=0$. This shows the transition is between a stable point and a stable limit cycle. When increasing $\omega$, the limit cycles increase radius to the point where the $I$-state exits the physical interval $[0,1]$. See Figure \ref{Fig:AllStability}.

We then investigate the effects of varying $\epsilon$ and $\mu$ for both the stable and unstable configurations. It turns out that the important transitions here happen at $\mu=0$ and $\epsilon=0$. See Appendix \ref{Sec:Stability_epmu}.

\begin{figure}[t]
	\centering
	\includegraphics[width=0.7\linewidth]{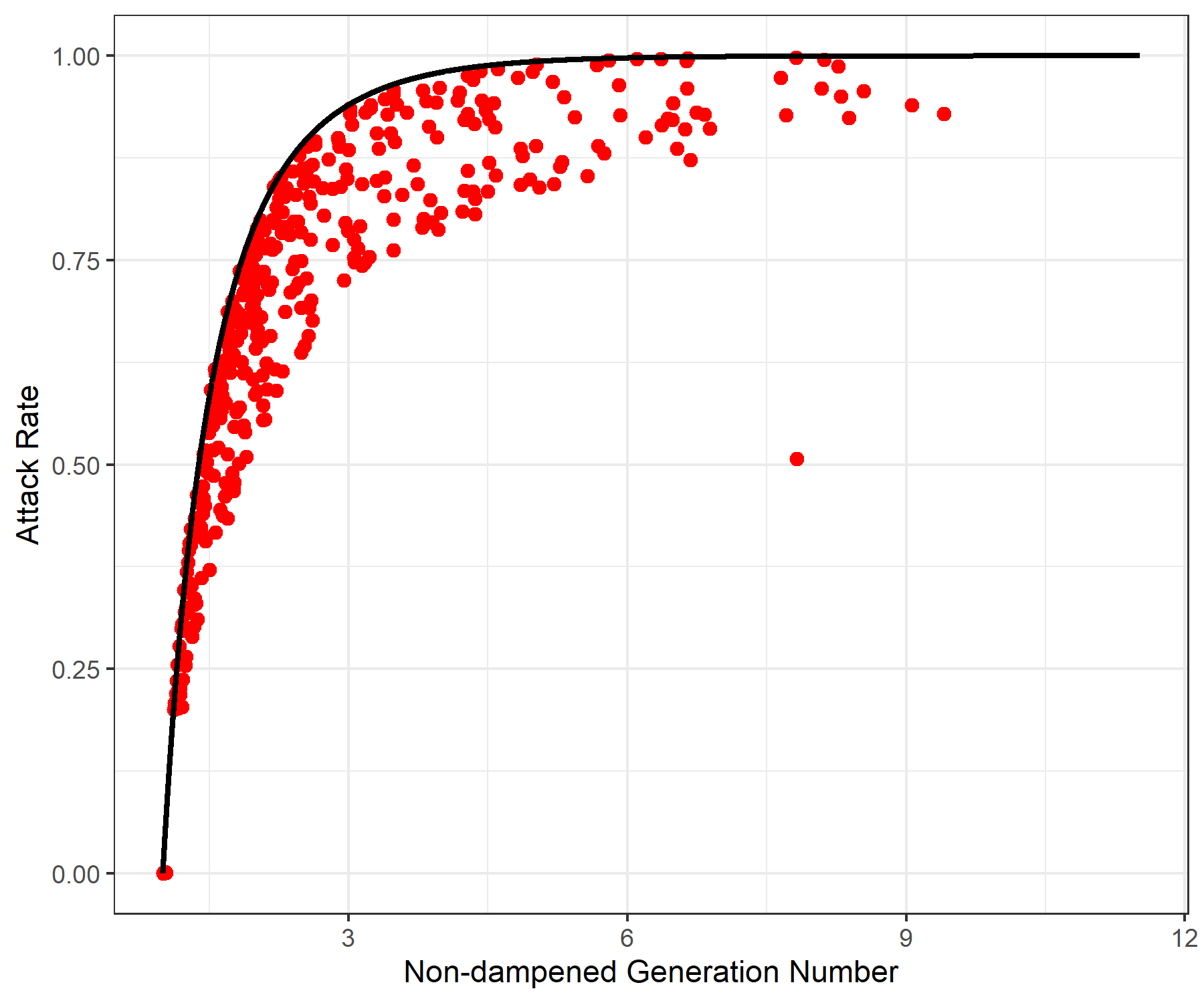}
	\caption{Connection between the initial non-damped reproduction number $R_{ND, 0}$ and the attack rate $R(\infty)$ for the kernel (\ref{Eq:Kernel}) using $n=1$. The black line indicates $R_{ND, 0} = -\frac{\ln\left(1 - R(\infty)\right)}{R(\infty)}$, which is the relation for no dampening. The red point are generated by uniformly sampling parameters in the intervals $\eta,\gamma\in [0.1,1]$, $\beta_0 \in  [\gamma, 2]$, and $c_0 \in [0.1,5], \sigma \in [0.1,1], \omega \in [0,1], \mu, \epsilon\in [-0.4,0.4]$, so the kernel is always positive. We see that all points lie to the right of the black line, indicating a lower attack rate than one would expect from a non-damped system ($c_0 = 0$).}
	\label{Fig:AttackRate}
\end{figure}

\subsection{Physiological Bounds on the Parameters}
In order to make realistic examples, let us briefly take a look at physiological scope of each parameter. Since we assume vanishing change in $S$ in the section, we must first and foremost require $S+E+I\leq 1\Rightarrow I^*(1+\frac{\gamma}{\eta})\leq 1-S$. This gives conditions on the feedback parameters through
\begin{eqnarray}\label{Eq:IstarCond}
    \frac{\beta_0 - \gamma/S}{c_0 \left(1+\epsilon\right) \left(\frac{1}{{\sigma}^2} + 2 \frac{\omega}{\sigma} \frac{\epsilon \left({\sigma}^2 - {\omega}^2\right) - 2\sigma\omega\mu}{\left({\sigma}^2 + {\omega}^2\right)^2}\right)} \leq \frac{S}{1 + \gamma/\eta}
\end{eqnarray}
for the general case and
\begin{eqnarray}
    \frac{{\sigma}^2\left(\beta_0 - \frac{\gamma}{S}\right)}{c_0} \leq \frac{S}{1 + \gamma/\eta}
\end{eqnarray}
for $\epsilon=\mu=\omega=0$. These are necessary conditions, but unfortunately not sufficient ones if the trajectories of $E$ and $I$ go above $1-S$ on their way to the equilibrium point. This may even be the case if $I(0)<I^*$, see Figure \ref{Fig:AllStability}.

Note that $\gamma$ and $\eta$ are typically equilibrium by the nature of the disease, though $\gamma$ can be artificially lowered by testing the population and putting positive cases in isolation. In terms of physical size, both $\sigma$ and $\omega$ have units of inverse time and it is unrealistic to have feedback that reacts faster than a day, unless the population have access to their own tests and report the results immediately (and accurately). We therefore consider $\sigma,\omega < 1\ { \rm days}\inv$.

We also require $\beta\geq 0$ at all times. A rough estimate can be obtained through Equations (\ref{Eq:betaInt}) and (\ref{Eq:Kernel}), where we for simplicity assume $\epsilon=\mu=\omega=0$.
\begin{eqnarray}\label{Eq:betaCond}
    \int\limits^t_{-\infty} \alpha(t-\tau)I(\tau)d\tau\ \leq\ (1-S) \int\limits^t_{-\infty} \alpha(t-\tau)d\tau\ = \nonumber \\ \ (1-S) \int\limits^{\infty}_{0} \alpha(u)du\ =\ (1-S) \frac{c_0}{\sigma^2}\ \leq\ \beta_0
\end{eqnarray}
If the equations (\ref{Eq:SEIR-model-Full}) are implemented as they stand, it is difficult to satisfy Equations (\ref{Eq:IstarCond}) and (\ref{Eq:betaCond}) at the same time, because we only control $c_0$ and $\sigma$. However, as non-pharmaceutical interventions hardly are implemented to satisfy Equation (\ref{Eq:betaCond}), we replace $\beta$ by $\max(0, \beta)$ by hand on the RHS of Equations (\ref{Eq:SEIR-model-Full}). The numerics show that the trajectory still converges to either a stable limit cycle or a stable equilibrium point, see Figure \ref{Fig:AllStability}.

\subsection{Attack Rate}
A different interesting question to ask is the effects of the feedback on the attack rate, i.e. $R(\infty)$. Note that at the end of the epidemic, the $I$-state is almost empty, and it is therefore tempting to think that the feedback does not affect the attack rate. This turns out not to be the case, as can be seen from the following. Start with the fraction of the $S$- and $R$-equations in the non-stratified model
\begin{eqnarray}
	\frac{dS}{dR} &=& -\frac{S \beta}{\gamma}\ =\ -\frac{S \beta_0}{\gamma} + \frac{S}{\gamma} \int\limits_{-\infty}^{t} \alpha(t-\tau) I(\tau) d\tau\nn\\
	\int\frac{dS}{S} &=& - \int\frac{\beta_0}{\gamma}dR + \frac{1}{\gamma} \int \int\limits_{-\infty}^{t} \alpha(t-\tau) I(\tau) d\tau dR\nn\\
	\ln\left(\frac{S(\infty)}{S(-\infty)}\right) &=& -\frac{\beta_0}{\gamma}\left(R(\infty) - R(-\infty)\right) - \frac{1}{\gamma} \int \int\limits_{-\infty}^{t} \alpha(t-\tau) I(\tau) d\tau dR
\end{eqnarray}
If we assume $S(-\infty) \approx 1$ and $R(-\infty) \approx 0$, the first terms simplify. Reapplying $dR = \gamma I dt$ and shifting the $\tau$-integral, we can rewrite the last term as
\begin{eqnarray}
	\ln\left(1 - R(\infty)\right) &=& -\frac{\beta_0}{\gamma}R(\infty) - \int\limits_{-\infty}^{\infty} \int\limits_{-\infty}^{0} I(t) \alpha(-\tau) I(\tau + t) d\tau dt\nn\\
	&=&  -R_{ND, 0}R(\infty) - \int\limits_{-\infty}^{\infty} \int\limits_{0}^{\infty} I(t) \alpha(\tau) I(t - \tau) d\tau dt\ ,\label{Eq:AttackRateFinal}
\end{eqnarray}
where $R_{ND, 0}$ denotes the initial non-damped reproduction number. (As $\beta(t = 0) = \beta_0$, we could also just call this the initial reproduction number.)
While the second term is a complicated integral to handle in general, we can draw some conclusions from it. As long as $\alpha(\tau)\geq 0$, which is the case for our kernel when $\epsilon\cos\left(\arctan(\mu/\epsilon)\right) + \mu\sin\left(\arctan(\mu/\epsilon)\right) \geq -1$, we can be certain that the integrand is non-negative too. This means that attack rate for a given initial reproduction number will be lower than one would expect for a standard SEIR-model. See Figure \ref{Fig:AttackRate} for a numerical check of this. A more general, but less transparent calculation for a stratified model ($n>1$) can be found in Appendix \ref{Sec:AttackRateApp}.

\section{Physiological Interpretation of Memory Effects}
While it is clear that Non-pharmaceutical interventions as well as a population reading about high rates of infection in the news and therefore changing their behavior are obvious examples of memory effects, we would like to briefly discuss the physiological interpretation in more details.\\
\newline
If we start with $\omega=0$, we only have a dampening from the kernel in Equation \eqref{Eq:Kernel}. So whenever the recent number of infectious has been high, the activity decreases.
The behavioral part is not limited to individual choice. Governmental non-pharmaceutical interventions, such as local lockdowns or extra tests in areas with high rates of infection, of course restrict activity based on recent history of infection, but it may also come from increased focus in the media leading to more cautious behavior. It may also come naturally, i.e.\ without testing the population, from the acquired immunity. If the option of moving from $S$ to $R$ is included, a lingering but waning immunity would take the same kernel form, where infectiousness diminishes for some time.
Note that self-isolation does not take this form as it is immediate in time, i.e.\ the individual isolates based on their own illness, not the previous illness of the population.\\
\newline
The interpretation of $\omega>0$ should be seen more exclusively as behavioral. It allows for a momentary increase in activity based on a previous wave of infection. Unless the disease weakens individuals and makes secondary infection more likely, this will not happen naturally. However, after a period of either lockdown or self-isolation, humans may be inclined to compensate socially, and thus a wave of infection may be followed by first a period of lower activity and then one of higher.
In other words, as we are more social than logical beings, an oscillatory kernel may be very relevant for description of adaptive behavior.\\
\newline
Transmission of a disease may also vary over one week due to different social behaviour in workdays as opposed to weekends. In this case the period of the oscillating disease transmission is 7 days.

\section{Conclusion}

We have implemented a feedback mechanism in epidemic models and illustrated its properties. This is useful for modelling adaptive behaviour and non-pharmaceutical interventions, which are both central in epidemic control.

The feedback dampening cannot prevent an outbreak from happening, as the stability of the disease-free equilibrium point does not depend on the kernel. Instead, the feedback can create equilibrium points locally in time as well as decrease the attack rate, so the severity of the outbreak can be contained.

The existence of a stable equilibrium point on short timescales is noteworthy, because it explains how the effective generation number of many countries during the COVID19-pandemic stayed consistently close to 1, which is impossible in normal SIR-type models without fine-tuning.

We have also shown that the feedback affects the attack rate $R(\infty)$. This is important, because it shows that non-pharmaceutical interventions reduce the number of people that need to be infected on long time scales, instead of simply postponing the time of infection, which would otherwise be reasonable to assume.

A common feature of the modelling frameworks presented in Section \ref{Sec:SEIR} is the presence of oscillatory time histories. We have generalized the methodology used in these sections to a linear chain trick methodology for general dynamical delay system with oscillations in its history. For the sake of completeness we have also shown in detail the continuous dependence of the solutions of the dynamical systems in the time histories. One important outcome of this analysis is the robustness property of the solutions. See Appendix \ref{Sec:AttackRateApp}.

\section*{Acknowledgements} The present work was initiated in September 2022 when J. Wyller (JW) was a Guest Researcher at Department of Applied
Mathematics and Computer Science, Technical University of Denmark. JW would like to express his sincere gratitude to Technical University of Denmark for kind hospitality during the stay. JW gratefully acknowledges the financial support from internal funding scheme at Norwegian University of Life Sciences (project number 1211130114), which financed the international stay at Department of Applied Mathematics and Computer Science, Technical University of Denmark, Denmark. AM is funded by Statens Serum Institut, Denmark. We are thankful for the eminent and profound contributions from Snif Mielke.



\bibliography{main_v2}


\begin{appendices}

\begin{sidewaysfigure}
	\centering
	\hspace{-40pt}\includegraphics[width=0.9\linewidth]{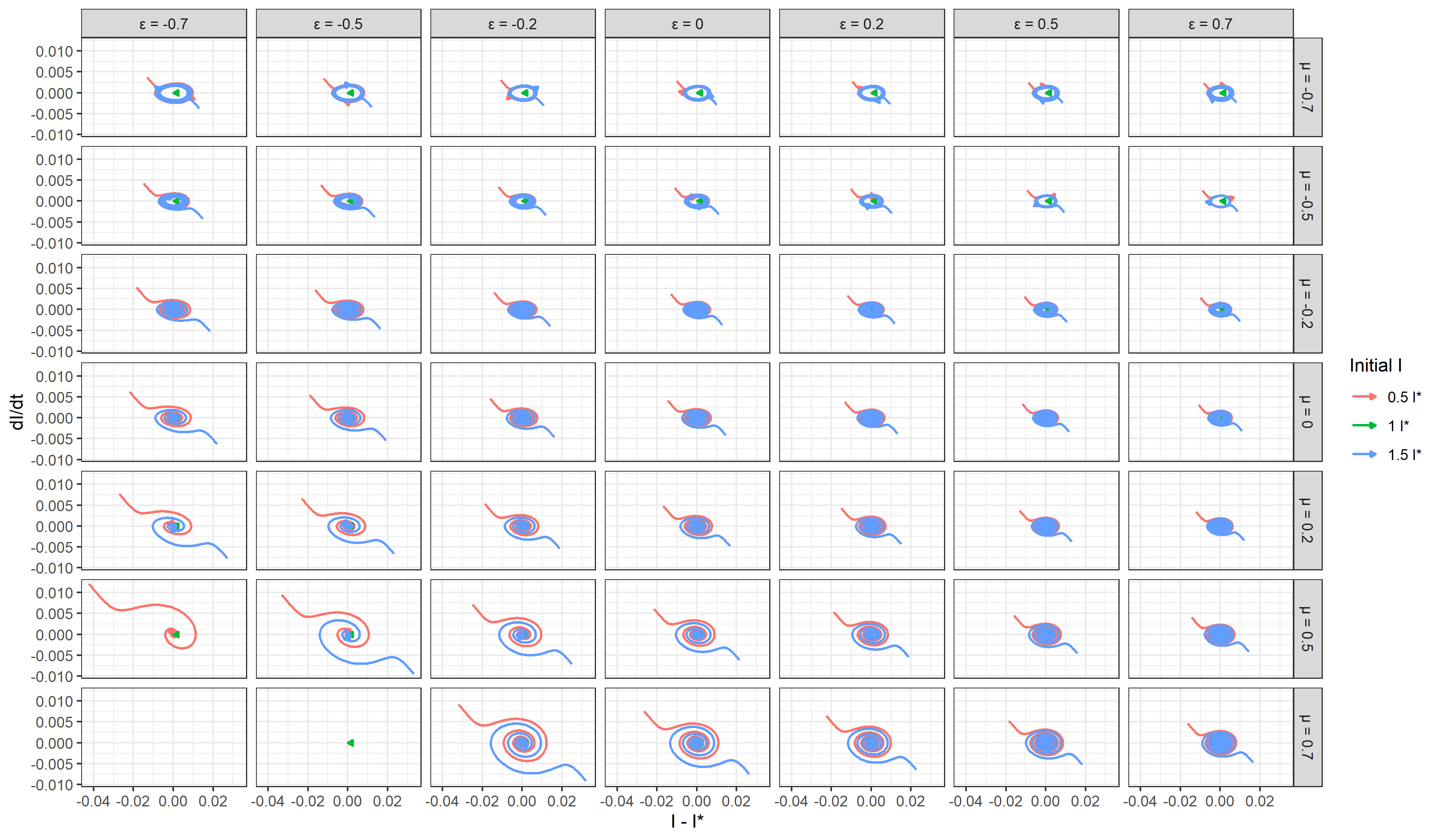}
	\caption{Phase space and stability for $\epsilon$ and $\mu$ using $n=1$. We keep $S$ constant to illustrate the local stability in time. The parameters are $\gamma = \eta = 1/3.5,\ c_0 = 5.5\cdot(1+0.5),\ \sigma = 0.5,\ S = 0.8,\ \beta=1.319$ for all the configurations, i.e., the top panel of the second column in Figure \ref{Fig:AllStability} with changing $\epsilon$ and $\mu$.
	The initial conditions are chosen according to the equilibrium point from Equation (\ref{Eq:FixedPoint}), but with $I$ varied to show convergence of different paths. Runs that exit the unphysical interval $E+I\in[0,1]$ are omitted. We have also rescaled $c_0\to\frac{c_0}{1+\epsilon}$ to remove trivial effects from the normalization.
	Here, the change in stability primarily occurs while going along the diagonal where $\epsilon$ and $\mu$ have opposite signs. We use $\max(\beta,0)$ instead of $\beta$ in the RHS of Equation \ref{Eq:SEIR-model-Full}}
	\label{Fig:AllStability_epmu_Stable}
\end{sidewaysfigure}

\begin{sidewaysfigure}
	\centering
	\hspace{-40pt}\includegraphics[width=0.9\linewidth]{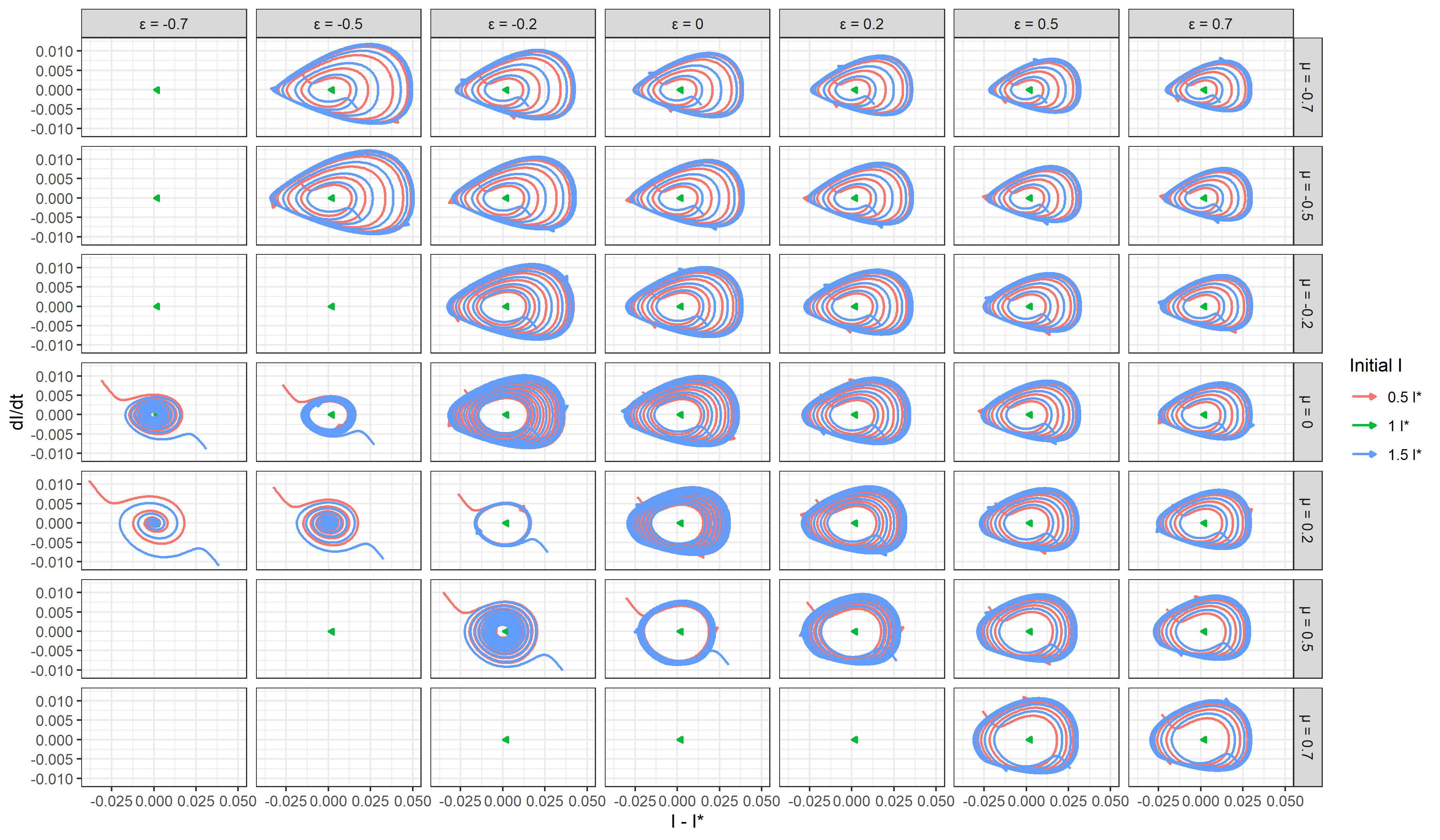}
	\caption{Phase space and stability for $\epsilon$ and $\mu$ using $n=1$. We keep $S$ constant to illustrate the local stability in time. The parameters are $\gamma = \eta = 1/3.5,\ c_0 = 5.5\cdot(1+0.5),\ \sigma = 0.5,\ S = 0.8,\ beta=1.719$ for all the configurations, i.e., the bottom panel of the second column in Figure \ref{Fig:AllStability} with changing $\epsilon$ and $\mu$.
	The initial conditions are chosen according to the equilibrium point from Equation (\ref{Eq:FixedPoint}), but with $I$ varied to show convergence of different paths. Runs that exit the unphysical interval $E+I\in[0,1]$ are omitted. We have also rescaled $c_0\to\frac{c_0}{1+\epsilon}$ to remove trivial effects from the normalization. This allows the model to cross the threshold where $I^*$ would otherwise diverge. 
	Here, the stability seems unaffected apart from for large positive values of $\mu$ and large negative values of $\epsilon$, where the solution eventually diverges. We use $\max(\beta,0)$ instead of $\beta$ in the RHS of Equation \ref{Eq:SEIR-model-Full}}
	\label{Fig:AllStability_epmu_Unstable}
\end{sidewaysfigure}

\section{Numerical Stability Analysis for Amplitude of Oscillations}\label{Sec:Stability_epmu}
The effects of varying $\epsilon$ and $\mu$ for the kernel (\ref{Eq:Kernel}) depend on whether a stable or unstable point is considered for $\omega=0$. Heuristically, it seems that the oscillations are able to break the stability of the equilibrium point, but not introduce it. It also seems that period doubling is possible for sufficiently large negative . See Figure \ref{Fig:AllStability_epmu_Stable} for a stable configuration and Figure \ref{Fig:AllStability_epmu_Unstable} for an unstable one.

\section{Stratified Attack Rate}\label{Sec:AttackRateApp}
	When extending to a stratified model, we must first decide what we mean by an attack rate. In the following, we will think of it as a vector. That is, the attack rate in each group is considered separately, so moving recovered from one group to another constitutes a different attack rate, even if the sum of infected is invariant.\\
	\newline
	Let us start with a time-independent $\beta$, i.e.\ with no memory effects. The integral version of a general model is \footnote{We assume we are not the first to derive the attack rate for a stratified model, but as we have not found it in the literature, we include the calculation here.}
	\begin{eqnarray}\label{Eq:Strat:Setup}
		\dot{S}_j(t)\ :=\ -C_j(t)\ =\ -S_j(t)\sum_k \beta_{jk}\int\limits_{-\infty}^{t}C_k(\tau)n_k(t-\tau)d\tau\ .
	\end{eqnarray}
	Here $C_j(t)$ is the number of new infections in group $j$ at time $t$ and $n_k(t)$ is the infection curve for a single individual in group $k$. We in this sense make very few assumptions about the model other than that $C$ has to be local in time. Note that we have explicitly separated $\beta$ from $n(t)$, such that $n(t)$ can remain the same over the course of the epidemic, even when considering time-dependent $\beta$ later.\\
	\newline
	We move $S_j(t)$ to the other side and integrate over $t$
	\begin{eqnarray}
		\log\bigg[\frac{S_j(\infty)}{S_j(-\infty)}\bigg] &=&  - \sum_k \beta_{jk}\int\limits_{-\infty}^{\infty}\int\limits_{-\infty}^{t}C_k(\tau)n_k(t-\tau)d\tau dt\ .
	\end{eqnarray}
	We can decouple the integrals
	\begin{eqnarray}\label{Eq:AttackRateDecoupling}
		\log\bigg[\frac{S_j(\infty)}{S_j(-\infty)}\bigg] &=& -\sum_k \beta_{jk}\int\limits_{-\infty}^{\infty}C_k(t) dt \int\limits_{0}^{\infty}n_k(\tau)d\tau\ .
	\end{eqnarray}
	We identify $\int\limits_{-\infty}^{\infty}C_k(t) dt = S_k(\infty) - S_k(-\infty)$ and the next-generation matrix $A_{jk} = \int\limits_{0}^{\infty} \beta_{jk} n_k(t) dt$. Rewriting $S_j(-\infty) \approx S_j(\infty) + R_j(\infty) = \nu_j$, and we are left with
	\begin{eqnarray}\label{Eq:Strat:Result}
		\log\left(1 - \frac{R_j(\infty)}{\nu_j}\right) &=& - \sum_k A_{jk} R_k(\infty)\ .
	\end{eqnarray}
	So having the same next-generation matrix is definitely a sufficient condition for the same attack rate, but any $A$ that gives the same solution to the Equation \eqref{Eq:Strat:Result} will give the same attack rate. In a sense, $R(\infty)$ is a kind of pseudo-eigenvector of $A$, though this is of course a much more difficult problem because of the non-linearity. (Note that expansion of the logarithm to obtain a linear system is non-nonsensical. As $R_j(\infty)$ is typically close to $\nu_j$, we would be expanding around a singularity.)\\
	\newline
	When adding time-dependence of the kind \eqref{Eq:betaInt}, we get
	\begin{eqnarray}\label{Eq:Strat:Setup}
		\dot{S}_j(t)\ &&:=\ -C_j(t)=-S_j(t)\sum_k (\beta_0)_{jk}\int\limits_{-\infty}^{t}C_k(\tau)n_k(t-\tau)d\tau\nn\\
		&&- S_j(t)\sum_k\sum_m \int\limits_{-\infty}^{t}C_k(\tau)n_k(t-\tau) \int\limits_{-\infty}^{\tau} (\alpha)_{jkm}(\tau-\tau') I_m(\tau') d\tau' d\tau\ .
	\end{eqnarray}
	We make the same rewriting while integrating over $t$
	\begin{eqnarray}
		&&\log\bigg[\frac{S_j(\infty)}{S_j(-\infty)}\bigg] = -\sum_k (\beta_0)_{jk} \int\limits_{-\infty}^{\infty}\int\limits_{-\infty}^{t}C_k(\tau)n_k(t-\tau)d\tau dt\nn\\
		&&- \sum_k\sum_m \int\limits_{-\infty}^{\infty} \int\limits_{-\infty}^{t}C_k(\tau)n_k(t-\tau) \int\limits_{-\infty}^{\tau} (\alpha)_{jkm}(\tau-\tau') I_m(\tau') d\tau' d\tau dt\ .
	\end{eqnarray}
	The time-independent part becomes the same as in Equation \eqref{Eq:AttackRateDecoupling}
	\begin{eqnarray}
		&&\log\left(1 - \frac{R_j(\infty)}{\nu_j}\right) = - \sum_k A_{jk}^{(0)} R_k(\infty)\nn\\
		&&- \sum_k\sum_m \int\limits_{-\infty}^{\infty} \int\limits_{-\infty}^{t}C_k(\tau)n_k(t-\tau) \int\limits_{-\infty}^{\tau} (\alpha)_{jkm}(\tau-\tau') I_m(\tau') d\tau' d\tau dt\ .
	\end{eqnarray}
	where $A_{jk}^{(0)} = \int\limits_{0}^{\infty} (\beta_0)_{jk} n_k(t) dt$ is the non-damped next generation matrix. As in Equation \eqref{Eq:AttackRateFinal}, the integral is not possible to solve in general, but it is strictly non-negative for $\alpha(t)\geq 0$. The equivalent argument is a lot less transparent because of the complicated nature of the relation \eqref{Eq:Strat:Result}, but the principle seems analogous, and we would therefore expect the attack rate to decrease for non-zero feedback.

\end{appendices}
\end{document}